%% file: main.tex
\documentclass[sigconf]{acmart}

\AtBeginDocument{%
  \providecommand\BibTeX{{%
    \normalfont B\kern-0.5em{\scshape i\kern-0.25em b}\kern-0.8em\TeX}}}

\copyrightyear{2021} 
\acmYear{2021} 
\setcopyright{rightsretained} 
\acmConference[CUI '21]{CUI 2021 - 3rd Conference on Conversational User Interfaces}{July 27--29, 2021}{Bilbao (online), AA, Spain}
\acmBooktitle{CUI 2021 - 3rd Conference on Conversational User Interfaces (CUI '21), July 27--29, 2021, Bilbao (online), AA, Spain}
\acmDOI{10.1145/3469595.3469626}
\acmISBN{978-1-4503-8998-3/21/07}

\usepackage{balance}
\usepackage{caption,subcaption}
\usepackage{graphics}      
\usepackage{color}
\usepackage{todonotes}
\usepackage{xspace}
\usepackage{soul}
\usepackage{makecell}
\usepackage[flushleft]{threeparttable}
\usepackage{xspace}
\usepackage{tikz}
\usepackage{xcolor}
\usepackage{color, colortbl, soul}
\usepackage[bottom]{footmisc}
\usepackage{dblfloatfix}

\def\S{Sec.\xspace}
\def\ie{\textit{i.e.,}\xspace}
\def\etal{\textit{et al.}\xspace}
\def\etc{\textit{etc.}\xspace}
\def\eg{\textit{e.g.,}\xspace}

\def\aka{\textit{a.k.a.,}\xspace}

\newcommand{\code}[1]{\texttt{#1}}
\newcommand{\red}[1]{\textcolor{black}{#1}}

\usepackage{csquotes}
\renewcommand{\mkbegdispquote}[2]{\itshape}

\begin{document}

% Toward configurable 
\title[Toward a Unified Metadata Schema for EMA with Voice--First Virtual Assistant]{Toward a Unified Metadata Schema for Ecological Momentary Assessment with Voice-First Virtual Assistants}

\author{Chen Chen}
\orcid{0000-0001-7179-0861}
\email{chenchen@ucsd.edu}
\affiliation{%
  \department{Computer Science and Engineering}
  \institution{University of California San Diego}
  \city{La Jolla}
  \state{CA}
  \country{USA}
}

\author{Khalil Mirini}
\email{khalil@ucsd.edu}
\affiliation{%
  \department{Computer Science and Engineering}
  \institution{University of California San Diego}
   \city{La Jolla}
  \state{CA}
  \country{USA}
}

\author{Kemeberly Charles}
\email{kecharle@ucsd.edu}
\affiliation{%
  \department{School of Medicine}
  \institution{University of California San Diego}
  \city{La Jolla}
  \state{CA}
  \country{USA}
}

\author{Ella T. Lifset}
\orcid{0000-0003-2640-3206}
\email{etlifset@ucsd.edu}
\affiliation{%
  \department{Biological Sciences}
  \institution{University of California San Diego}
  \city{La Jolla}
  \state{CA}
  \country{USA}
}

\author{Michael Hogarth}
\orcid{0000-0002-4264-1258}
\email{mihogarth@ucsd.edu}
\affiliation{%
  \department{School of Medicine}
  \institution{University of California San Diego}
  \city{La Jolla}
  \state{CA}
  \country{USA}
}

\author{Alison A. Moore}
\orcid{0000-0003-2989-4346}
\email{alm123@ucsd.edu}
\affiliation{%
  \department{School of Medicine}
  \institution{University of California San Diego}
  \city{La Jolla}
  \state{CA}
  \country{USA}
}

\author{Nadir Weibel}
\orcid{0000-0002-3457-4227}
\email{weibel@ucsd.edu}
\affiliation{%
  \department{Computer Science and Engineering}
  \institution{University of California San Diego}
  \city{La Jolla}
  \state{CA}
  \country{USA}
}

\author{Emilia Farcas}
\orcid{0000-0001-6485-0141}
\email{efarcas@ucsd.edu}
\affiliation{%
  \department{Qualcomm Institute}
  \institution{University of California San Diego}
  \city{La Jolla}
  \state{CA}
  \country{USA}\\[.8cm]
}

\renewcommand{\shortauthors}{C. Chen, K. Mirini, K. Charles, E.T. Lifset, M. Hogarth, A.A. Moore, N. Weibel, E. Farcas}

\input{00-abstract}


\maketitle
\input{01-introduction}

\input{02-design}
\input{03-implementation}

\input{04-conclusions}
\input{05-ack}

\balance
\bibliographystyle{ACM-Reference-Format}
\bibliography{references}

\end{document}

%% file: 00-abstract.tex
\begin{abstract}
Ecological momentary assessment~(EMA) is used to evaluate subjects' behaviors and moods in their natural environments, yet collecting real-time and self-report data with EMA is challenging due to user burden.
Integrating voice into EMA data collection platforms through today's intelligent virtual assistants~(IVAs) is promising due to hands-free and eye-free nature.
However, efficiently managing conversations and EMAs is non-trivial and time consuming due to the ambiguity of the voice input.
We approach this problem by rethinking the data modeling of EMA questions and what is needed to deploy them on voice-first user interfaces.
We propose a unified metadata schema that models EMA questions and the necessary attributes to effectively and efficiently integrate voice as a new EMA modality.
Our schema allows user experience researchers to write simple rules that can be rendered at run-time, instead of having to edit the source code.
We showcase an example EMA survey implemented with our schema, which can run on multiple voice-only and voice-first devices.
We believe that our work will accelerate the iterative prototyping and design process of real-world voice-based EMA data collection platforms. 
\end{abstract}

%% file: 01-introduction.tex
\section{introduction}~\label{sec::intro}
%
% collecting data of EMA is important, cite some uEMA paper
Ecological momentary assessment (EMA) is an important technique in behavioral science to collect \textit{in situ} research participants' behaviors, experiences, and moods in their natural setting~\cite{Stone1994}.
Collecting real-time and temporally-dense participants' self-report EMA data in an ecologically valid setting is valuable, yet challenging, especially for under-represented groups (\eg~older adults and people with physical or mental disabilities)~\cite{Nagel2004, Ferreira2014}. 
The design of EMA data collection platforms must consider various interconnected concerns, such as user engagement, reporting burden, data validity, and honest disclosure~\cite{Doherty2020}.

Over the last few years, smartwatches have been considered as an effective tool to support EMA. 
For example, Intille~\etal~\cite{Intille2016uEMA, Ponnada2017, Ponnada2021} designed the $\mu$EMA on smartwatch using a 1-tap glance-able microinteraction to collect participants' mood states to understand the instantaneous mood that participants felt while getting notifications.
They found that $\mu$EMA can reduce the perceived cognitive burden and device access time, and thus increase the EMA response rate~\cite{Intille2016uEMA, Ponnada2017}.

Integrating \emph{voice interfaces} into EMA platforms also promises to increase user engagement, as participants would benefit from the hands-free and eye-free nature of voice-based devices. Using those devices can also reduce the device access time (\ie~time that participants take to access the device and start the EMA questionnaire) and usage time~(\ie~time for participants to complete the assigned list of EMA questions).
Researchers also proposed the \emph{voice-first} design~\cite{Bajorek2018} attempting to incorporate an additional visual modality allowing users to interact through a built-in touch screen. 
This shift is also visible in today's intelligent virtual assistants (IVAs) that have been built into various voice-enabled smart devices and often include additional displays and touch interfaces (\eg~Echo Show\footnote{Amazon Echo Show: \url{https://www.amazon.com/Echo-Show-8/dp/B07PF1Y28C}}). 
%

% the facing population is researchers and practitioners
Designing a usable conversational system for EMA data collection on such platforms is challenging.
When using methods such as rapid prototyping and iterative design, Human-Computer Interaction~(HCI) and User Experience (UX) researchers need to carefully consider how to design both the information output (\eg~how to announce EMA questions in a correct form and prompt users upon failures?) and information input (\eg~what are the users' possible intents?)~\cite{Holland2021, Sayago2019}.
Prototyping such EMA systems on IVAs, while carefully considering those questions, is non-trivial and time consuming, and currently hinders the design process.
To better understand the hurdles of deploying EMAs on voice-based IVAs, we break down the design process into three core challenges.

% why this is hard
\textbf{First}, current commercial systems are not designed with the concept of a prototype in mind. To evaluate the usability of a conversational voice user interface, complete systems need to be developed in complex IVA environments geared to develop and deploy products, which often means losing control on the application itself. 
To maintain more control, existing works attempted to build hardware-software systems from scratch, such as speech systems using a Raspberry Pi board for conducting Wizard of Oz studies~\cite{Brggemeier2019}. 
However, it is difficult for these custom-based solutions to scale to large research projects, and it is not feasible to replicate multiple such prototypes for large populations. 
In addition, the recording of private speech (\ie~conversations outside of the EMA research questions) on the vendor's cloud infrastructure is  problematic, as it could potentially break multiple ethical research regulations.
To address this privacy issue, MicShield~\cite{Sun2020} proposed using an additional mechanism to ensure that the privacy of unintended speech is preserved, using inaudible ultrasound signals that mask conversations not directed to the voice-based system. 
However, such approaches would incur additional prototyping time and effort.
Similar to the example above, these cutting-edge research approaches are far from being integratable in systems that can actually be deployed with research subjects, leading to issues of scalability and accessibility.

\textbf{Second}, although commercially-available voice survey platforms (\eg~SurveyLine\footnote{SurveyLine: \url{https://www.surveysbyvoice.com}}) provide a conversational speech system for potentially collecting EMA data, such platforms only offer standard services that are difficult to customize to the needs of a particular study. For example, existing platforms do not allow researchers to configure the occurrence of EMA questions based on context (\eg~time of the day) in an easy and flexible way. Furthermore, depending on the type of data being collected, studies may be subject to local data privacy or health laws such as the Health Insurance Portability and Accountability Act (HIPAA) in the United States, which further limit the third-party services that can be used for research.

\textbf{Third}, a practical voice-based EMA data collection platform is beyond a simple sequential question and answering interface and it usually requires more advanced, yet essential features, such as conditional branching and combining input from different type of responses~\cite{Ziegler2018}.

Although commercially-available voice assistant vendors provide rule-based intent design approaches for developers to fast prototype voice apps using serverless functions, the stateless nature of these solutions makes it difficult to track conversation flows.
Graphical programming methods for defining the conversation flows (\eg~VoiceFlow\footnote{VoiceFlow: \url{https://www.voiceflow.com}} and kore.ai\footnote{Kore.ai: \url{https://kore.ai}}) are geared to solve this problem, but currently do not offer enough flexibility and have important scalability issues. 

In this work, we propose the design of a metadata schema and programming model to support healthcare and behavioral science researchers to rapidly prototype a practical EMA data collection system that can be easily provisioned on voice-first digital assistants. 
Instead of advocating for the design of a customized hardware-software system from scratch, we leverage commercially available hardware that is affordable and easy to set-up.

To the end, we implement an EMA data collection platform using our proposed metadata schema on top of Amazon-based voice-first digital assistants (Amazon Echo Dot, Amazon Echo Show, and Alexa assistant running on top of iPhone and Apple Watch) with Amazon DynamoDB as the back-end data storage.
Although we are using Alexa assistants as the running example, our proposed schema can be transferred and generalized to other commercially available voice-first digital assistants.

%% file: 02-design.tex
% \hl{TODO: change Patient User to "User"}
\begin{figure*}[t]
    \centering
    \includegraphics[width=.7\textwidth]{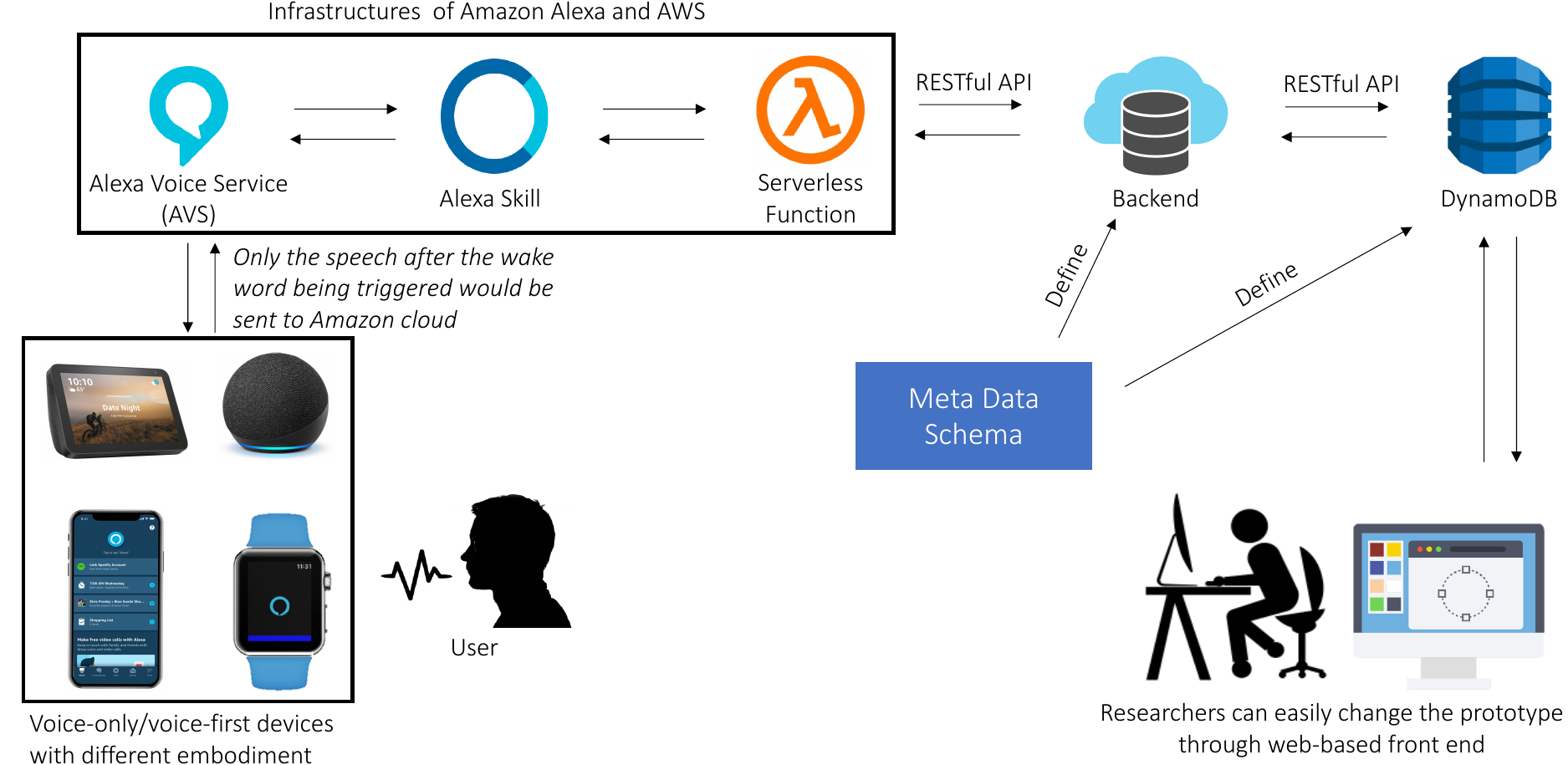}
    %\vspace{-0.5cm}
    \caption{Example system that uses our designed schema to store and render EMA questionnaires. To try out different conversation flows during the iterative design process, UX researchers only need to modify the content in the database.}
    \label{fig::architecture}
    \vspace{-0.3cm}
\end{figure*}

\begin{figure*}[b]
    \centering
    \includegraphics[page = 2, width=0.65\textwidth]{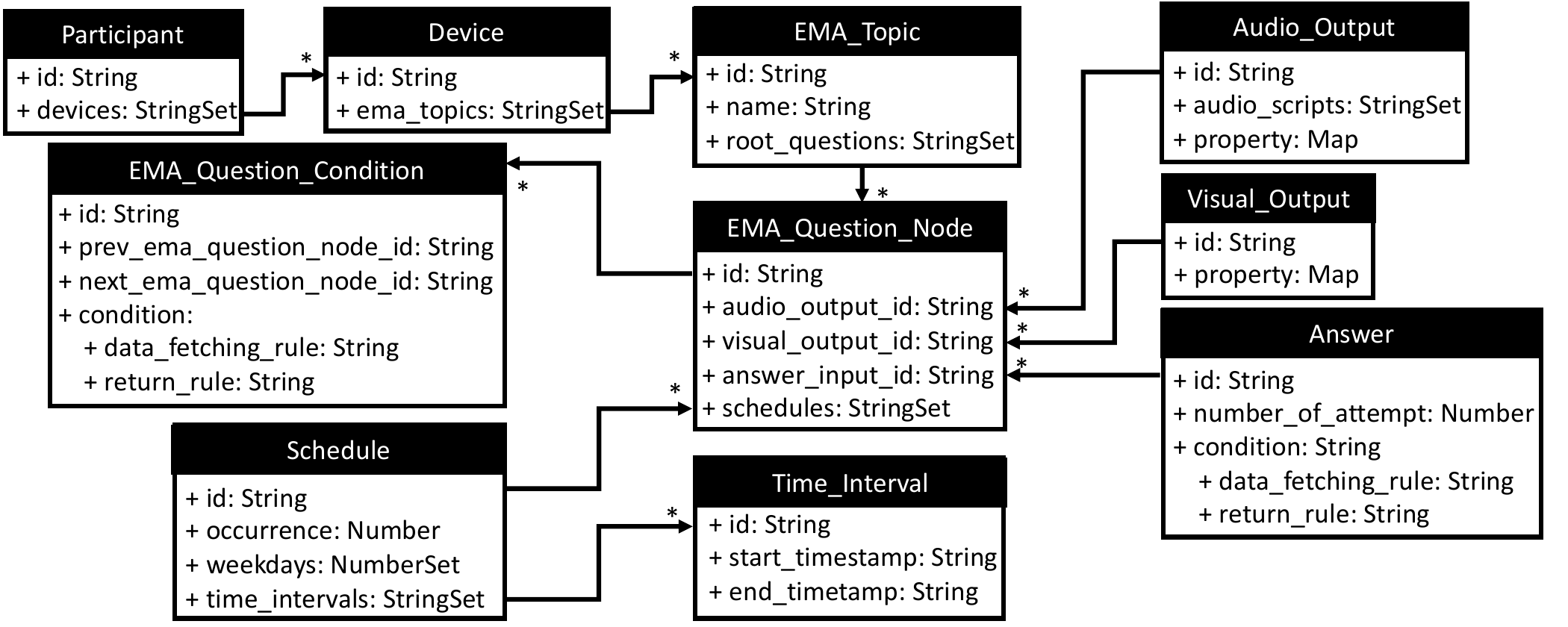}
    %\vspace{-0.5cm}
    \caption{A simplified entity relationships diagram for EMA questions and the necessary attributes. As a running example, we used the types supported by DynamoDB~\cite{TypeAmazonDB}. Notably, the \textit{property} fields in \textit{Audio\_Output} and  \textit{Visual\_Output} collections vary among types of questions.}
    \label{fig::erd}
    \vspace{-0.5cm}
\end{figure*}

\section{Schema 
Design}~\label{sec::sys}
With the Alexa ecosystem as an example, we consider an end-to-end system like the one shown in Figure~\ref{fig::architecture}, where the user's speech is captured through various devices with voice-based IVAs. 
The Alexa Voice Service~(AVS) is used to process and understand the user's speech.
An Alexa skill and lambda function are used to parse and forward the transcribed text, and to receive the EMA questions with graphic and/or audio elements to be rendered or announced through the front-end hardware.

Our proposed schema is used to define how a database stores the EMA questionnaires and how the back-end infrastructure manages the conversation flow.
Notably, during the iterative design process, researchers only need to change the meta attributes stored in the database and do not have to engage with source code.
We envision the creation of a web-based interface for researchers to specify the attributes.

Designing a schema to support prototyping EMAs on top of today's voice-first IVAs is non-trivial due to the diverse possibilities of users' intents, the requirement of controlling the occurrence of EMA questions based on ``real world'' contexts~\cite{Bajorek2018}, and the complexities of additional modalities provided by  voice-first interfaces (\eg~the visual output and touch input).
In this section, we describe the design and implementations of primitive building components.

%
% \hl{Change Audio to Audio\_Output and Visual to Visual\_Output}

\subsection{Entity Relationships and Cross-Platform Support}
Figure~\ref{fig::erd} shows the entity-relationships diagram of our proposed schema. We model each entity as a separate collection. Relations connect different entities of the voice-first EMA infrastructure.
One of our target groups are UX researchers who aim to rapidly prototype a voice-based EMA platform for evaluating usability, without needing to build state machines for managing conversation flows. Therefore, we designed our model to be minimal, intuitive, and flexible enough for them to tune and reconfigure the system. We now describe the main entities in our model:

\vspace{.5em}\noindent$\bullet$ 
\textbf{\texttt{EMA\char`_Topic}} and \textbf{\texttt{EMA\char`_Question\char`_Node}} encapsulate the EMA questions. Multiple \textbf{\texttt{root\char`_questions}} can be defined as part of \textbf{\texttt{EMA\char`_Topic}}. This is where the occurrence of each EMA question would be triggered based on the user-defined context. Also, multiple paraphrased questions can be provided for the same topic, and the system can pick randomly between them to reduce user boredom.

\vspace{.5em}\noindent$\bullet$
Since our schema is designed for different types of multi-modal devices, we introduce \textbf{\texttt{Visual\_Output}} and \textbf{\texttt{Audio\_Output}} as two separated collections shareable by multiple \textbf{\texttt{EMA\_Question\_Nodes}}.
     For example, multiple EMA questions with a $5$-point Likert scale input might share the same \textbf{\texttt{Visual\_Output}} and define five buttons when deployed on voice-first devices with touchable input.  
     Notably, multiple audio scripts can be pre-defined as part of the \textbf{\texttt{Audio\_Output}} entity, which aims to provide users a feeling of conversation, potentially enhancing the user engagement~\cite{Surveymonkey2021}. 
    The \textbf{\texttt{Visual\_Output}} collection defines basic properties for instructing voice-first devices to render the graphic interactive widgets (\eg~buttons, sliders, \etc). 

\vspace{.5em}\noindent$\bullet$
Our schema also defines the concept of \textbf{\texttt{Answer}}, which provides a way for practitioners to define the correct rule for validating users' responses and providing feedback to users (\eg~prompting error messages) (see \S\ref{sec::design::answer}).
   We specify the \textbf{\texttt{number\_of\_attempts}} as part of the \textbf{\texttt{Answer}} collection as the maximum number of times that participants may correct their previous responses. 

\vspace{.5em}\noindent$\bullet$
Each \textbf{\texttt{EMA\_Question\_Node}} can connect to a number of different \textbf{\texttt{EMA\_Question\_Conditions}}, indicating the possible transitions between conversations.
    Also, each \textbf{\texttt{Schedule}} can be shared by multiple \textbf{\texttt{EMA\_Question\_Nodes}}, where the occurrences of each question can be determined by the different time contexts.

\begin{figure*}[b]
     \centering
     \begin{subfigure}[b]{0.24\textwidth}
         \includegraphics[page=1, width=\textwidth]{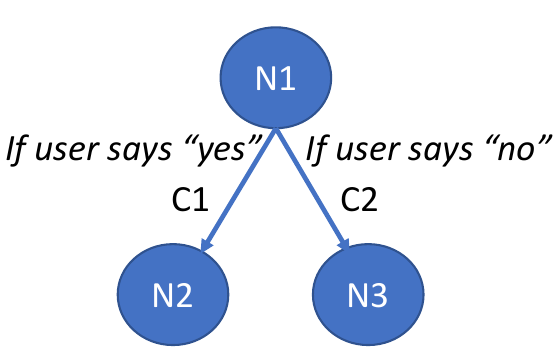}
         \caption{Decision tree modelling.}
         \label{fig::modeling_conditional_branching}
     \end{subfigure}
     %\hfill
     \begin{subfigure}[b]{0.24\textwidth}
         \centering
         \includegraphics[page=2, width=\textwidth]{figures/conditional-branching-erd.pdf}
         \caption{Example object of \code{C1}.}
         \label{fig::c1_example}
     \end{subfigure}
     %\hfill
     \begin{subfigure}[b]{0.24\textwidth}
         \centering
         \includegraphics[page=3, width=\textwidth]{figures/conditional-branching-erd.pdf}
         \caption{Example object of \code{C2}.}
         \label{fig::c2_example}
     \end{subfigure}
     %\hfill
     \begin{subfigure}[b]{0.24\textwidth}
         \centering
         \includegraphics[page=4, width=\textwidth]{figures/conditional-branching-erd.pdf}
         \caption{Example object with RPC.}
         \label{fig::rpc_example}
     \end{subfigure}
     %\vspace{-0.3cm}
     \caption{Modelling of conditional branching using a decision tree graph.}
     \label{fig::example_conditional_branching}
     \vspace{-0.5cm}
\end{figure*}

\subsection{Contextual Awareness and Conditions}~\label{sec::design::context}
Contextual awareness requires the occurrence of specific EMA questions, depending on predefined real-world conditions (\ie~the context).
For example, such context can include a particular time, the weather obtained from a remote weather services, previous answers, and sensor data.
Our model allows the flow of the EMA conversation to largely depend on such contextual information.
Due to the context being highly heterogeneous, our schema includes a method for researchers to easily define contextual rules during initial prototypes.

Inspired by the \emph{function} component available in many programming languages, we incorporated a \textbf{\texttt{condition}} property that includes both a \textbf{\texttt{data\_fetching\_rule}} and \textbf{\texttt{return\_rule}} as part of the \textbf{\texttt{EMA\_Question\_Condition}}.
The \textbf{\texttt{data\_fetching\_rule}} field contains code defining rules for either fetching remote data or processing input data; the \textbf{\texttt{return\_rule}} field contains code defining the condition to evaluate the input from the data fetching.
%string of code which would be passed back to our main server.

We now use \emph{conditional branching} as a running example to describe how our model supports contextual awareness, specifically how a participant's previous EMA response is used as the context to decide the next EMA question.
The use of conditional branching (\aka~skip logic) is a well-known practice in designing questionnaires, where the respondents receives a different question based on how they answer the current question. 
Existing work~\cite{ConditionalBranching} has shown the benefits of using the principle of micro-interaction, which aims to reduce device access-time and usage-time with the goal of decreasing completion time, dropout rate, and support more accurate data entry. It has been demonstrated that this method is effective specifically in increasing compliance rate and completion rate if applied to the design of EMA data collection platforms on smartwatches~\cite{ConditionalBranching}. We think that applying this approach also to voice-based IVAs, and thus grounding our model on it, would result in similar increased efficiency.

In particular, our schema models each EMA questionnaire as a decision tree, where questions and conditions are modelled as nodes and edges of the tree~(see Fig.~\ref{fig::modeling_conditional_branching}). The \textbf{\texttt{EMA\_Question\_Condition}} entity contains the ID of two connected \textbf{\texttt{EMA\_Question\_Nodes}}:  \textbf{\texttt{prev\_ema\_question\_node\_id}} and  \textbf{\texttt{next\_ema\_question\_node\_id}} (see Fig. \ref{fig::c1_example} and \ref{fig::c2_example}).
Figure~\ref{fig::example_conditional_branching} also illustrates how we model an example question when the response from the previous question is used as the context.
Notably, researchers can simply treat the \textbf{\texttt{\_answer\_}} property as a built-in variable that stores the response from users.
When the response is stored, then a rule defining the success or failure of a particular condition can be devised based on the user input, which can be therefore used to determine the subsequent EMA question based on the previous response.

%Figure~\ref{fig::example_answer_rpc} 
Figure~\ref{fig::rpc_example} shows another example where researchers can define a remote procedure call (RPC) to collect data (\eg~current temperature) from a third-party service.
At run time, we use \textbf{\texttt{fork()}} and \textbf{\texttt{exec()}} as techniques to spawn other processes to execute the rule defined, if this is required.
Our current prototype uses the spawned process as an additional sandbox to execute the rules. However, in the long term, alternative isolated sandbox (\eg~containers and remote instances) might be considered to address scalability and security challenges.

\subsection{Schedule and Occurrence}~\label{sec::design::schedule}
Often the EMA question should occur at a specific time, which can be defined by the researchers in our meta-model as well.
To do that, our schema realizes and natively integrates the concepts of \emph{schedule} and \emph{occurrence}.
The \textbf{\texttt{Schedule}} entity defines the range of time when each question is scheduled to be prompted each day.
In order to model the concept of occurrence, we take into consideration the timestamp of previous attempts of the same question, which is cached on the back-end.
%\hl{TODO: where is the timestamp stored in the model?}
%
For example, if we define \textbf{\texttt{occurrence}} as $3600$ seconds and \textbf{\texttt{max\_number\_of\_occurrence}} as $2$, this means that in any one hour interval, the same question can be prompted no more than twice.

\subsection{Answer Validation and Error Prompt}~\label{sec::design::answer}
Compared to approaches that use a Graphical User Interface~(GUI) with touch and mouse-click based inputs, a key challenge that voice-first devices are facing is the ambiguity and unpredictability of users' inputs~\cite{Cohen2004}.
This means that EMA systems based on voice-driven conversations need a way to provide explicit error-recovery guidance.
However, due to the variety of errors that can occur during the execution of voice-based applications, it is impractical to provide appropriate error messages at prototyping time.
This means that the design of error messages can only be achieved after multiple rounds of the iterative process. 
To address this problem, we included in our answer validation schema specific ways to define error prompting messages during the EMA testing phase. 

\begin{figure*}[t]
     \centering
     \begin{subfigure}[b]{0.49\textwidth}
         \centering
         \includegraphics[width=\textwidth]{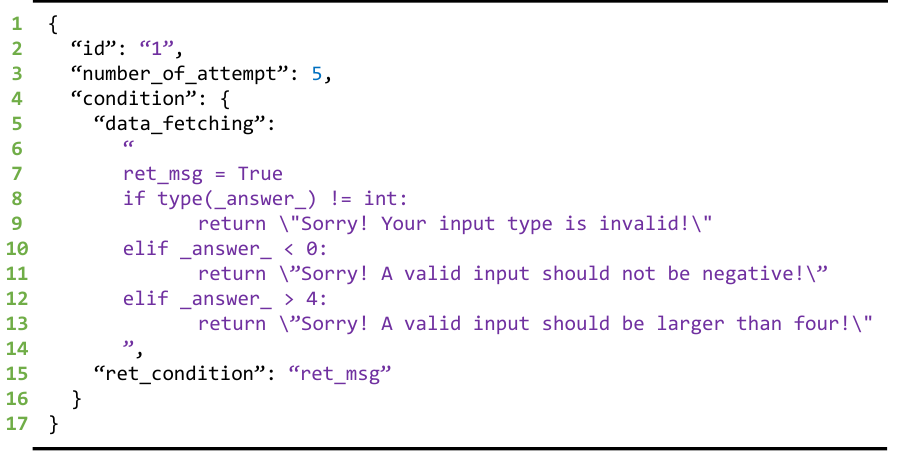}
         %\vspace{-0.2em}
         \caption{Example for validating 5--point Likert input.}
         \label{fig::example_answer_likert}
     \end{subfigure}
     %\hfill
     \,\,\,\,\,\,\,
     \begin{subfigure}[b]{0.30\textwidth}
         \centering
         \includegraphics[width=\textwidth]{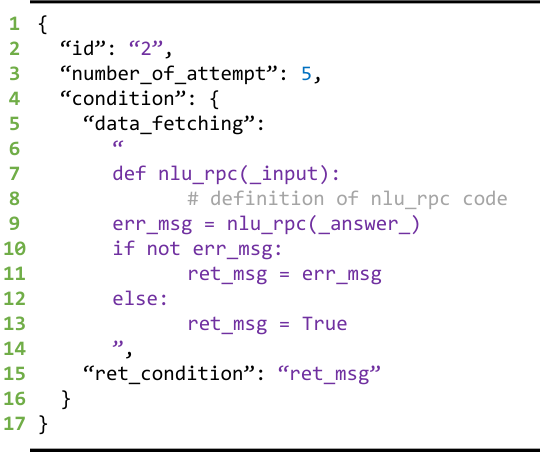}
         \caption{Example using remote cloud services.}
         \label{fig::example_answer_rpc}
     \end{subfigure}
     \vspace{-0.2cm}
     \caption{Example of how to instantiate an \code{Answer} in our metadata schema.}
     \label{fig::example_answer}
    % \vspace{-0.5cm}
\end{figure*}

Similar to context descriptions (see \S\ref{sec::design::context}), these error message rules are rendered at run time using \emph{reflection} techniques. Instead of requiring researchers to revisit the source code, only little effort is required to write these \red{rules}. 
Figure~\ref{fig::example_answer_likert} shows an example \textbf{\texttt{Answer}} entity and how it can be used to validate a typical 5-point Likert scale input.
%
%Notably, the practitioner will be asked to specify the \textbf{\texttt{number\_of\_attempt}} to prevent same questions being prompted infinitely.
% 
As part of future work, our schema could also support researchers to use RPC, where a remote cloud service such as \cite{witai} can check the correctness of participants' responses in real-time using smarter natural language understanding models, and thus generate corresponding error messages (Fig.~\ref{fig::example_answer_rpc}).

%% file: 03-implementation.tex
\begin{figure}[b]
    \vspace{-0.2cm}
    \centering
    \includegraphics[width=0.4\textwidth]{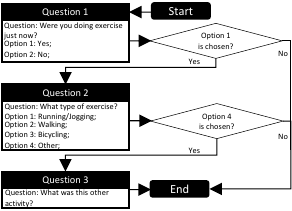}
    \vspace{-0.2cm}
    \caption{Example EMA survey for evaluating sedentary behavior and physical activity revised from \cite{Maher2018}.}
    \label{fig::example_ema_survey}
    \vspace{-0.2cm}
\end{figure}

\begin{figure*}[t]
    \vspace{-0.2cm}
\centering
    \includegraphics[width=1\textwidth]{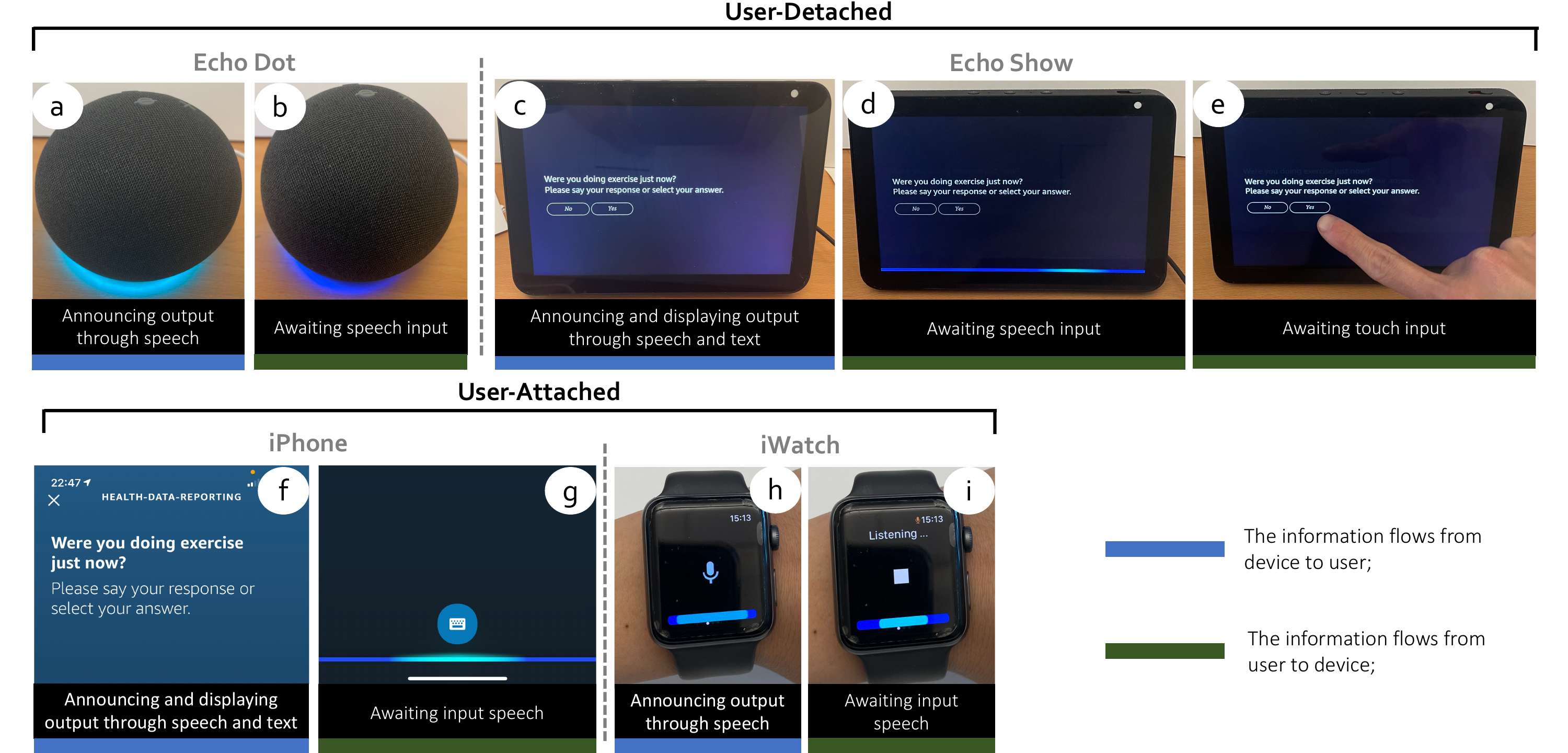}
    \caption{Example EMA applications implemented on the standalone voice-only user interface realized by Amazon Echo Dot (a -- b), the standalone voice-first user interface realized by Amazon Echo Show, where input can be achieved through speech (d) and touch (e), the voice-first interface realized by Amazon Alexa applications on iPhone (f -- g), and voice-only user interface on wrist wearables, realized by Apple iWatch (h -- i).}
    \label{fig::example_showcase}
\end{figure*}

\section{Example Applications}\label{sec::eval}
To better showcase our proposed schema we outline how to rapid prototype an example EMA questionnaire with conditional branching properties.
We reused the EMA questionnaire designed by Maher~\etal~\cite{Maher2018}, to evaluate the sedentary behavior and physical activity of older adults~(see Fig.~\ref{fig::example_ema_survey}) and deployed it on Amazon Alexa IVAs, using DynamoDB to implement our data model.
We built an end-to-end system as shown in Figure~\ref{fig::architecture} by integrating an Alexa skill, lambda functions, and a python flask server.\footnote{Python Flask: \url{https://flask.palletsprojects.com}} 
We used the Alexa Presentation Language (APL) to define the attributes of the visual elements~\cite{apl}.

Figure~\ref{fig::example_showcase} shows our example application deployed on different \textit{user-detached} voice-first devices---usually standalone and affiliated to a particular fixed environment---and \textit{user-attached} voice-first devices---typically carried or worn by users---using our proposed metadata schema.
Figure~\ref{fig::example_showcase}a and ~\ref{fig::example_showcase}b\ show the prototype running on Amazon Echo Dot, which is user-detached and only supports speech input.
Figure~\ref{fig::example_showcase}c,~\ref{fig::example_showcase}d~and~\ref{fig::example_showcase}e show the prototype running on Amazon Echo Show, a popular user-detached voice-first user interface with built-in touch screen.
With our schema, researchers were able to easily define a number of input widgets beyond voice, using touchable buttons (see Fig.~\ref{fig::example_showcase}e).
Similarly, our prototype can also run on user-attached mobile devices.
We showcase our example using the Alexa App running on iPhone 12 Pro (see Fig~\ref{fig::example_showcase}f~and~-~\ref{fig::example_showcase}g).
As the current Alexa app is not supported by the Apple Watch Series $3$, we use the \textit{Voice in a Can}\footnote{Voice in a Can: \url{https://voiceinacan.com}} framework to receive, process, and render the interaction process (see Figs~\ref{fig::example_showcase}h~and~\ref{fig::example_showcase}i). 
Notably, in order to modify the questions, answer types, visual and audio outputs, and the conversation flow, researchers only need to refine the rules and metadata in the database, instead of revisiting any of the source code. 

\red{As} this paper focuses on the design of a schema to assist rapid proof-of-concept prototypes, we leave the analysis of the effectiveness of different user interfaces and hardware embodiment for future work.

%% file: 04-conclusions.tex
\section{Conclusion}
We designed a novel unified metadata schema to enable Human-Computer Interaction (HCI), user experience (UX), and behavioral health researchers to rapidly prototype Ecological Momentary Assessment (EMA) data collection applications on different voice-first smart devices with built-in IVAs.
The proposed schema supports critical features to enable the deployment of effective EMA surveys, and addresses multiple considerations and challenges introduced by the voice modality.
We showcased an EMA implementation example that assess the physical activities and sedentary behaviors for older adults using user-attached and user-detached Amazon Alexa enabled devices.

We believe that our work will accelerate the design and prototyping process of voice-integrated EMA data collection platforms, and will ultimately enable voice-based IVAs to be used as a support for EMA data collections.

%% file: 05-ack.tex
\section{Acknowledgements}
This work is part of early effort of Project VOLI\footnote{Prof. Michael Hogarth has an equity interest in LifeLink Inc. and also serves on the company’s Scientific Advisory Board.  The terms of this arrangement have been reviewed and approved by the UC San Diego in accordance with its conflict of interest policies.}~\cite{voli, volisite, Mrini2021MedicalQU}, and was supported by NIH/NIA under grant R56AG067393.
We appreciate insightful feedback from the anonymous reviewers and fellow colleagues at UC San Diego.

%% file: main.bbl

\begin{thebibliography}{23}

%%% the Web address.

\ifx \showCODEN    \undefined \def \showCODEN     #1{\unskip}     \fi
\ifx \showDOI      \undefined \def \showDOI       #1{#1}\fi
\ifx \showISBNx    \undefined \def \showISBNx     #1{\unskip}     \fi
\ifx \showISBNxiii \undefined \def \showISBNxiii  #1{\unskip}     \fi
\ifx \showISSN     \undefined \def \showISSN      #1{\unskip}     \fi
\ifx \showLCCN     \undefined \def \showLCCN      #1{\unskip}     \fi
\ifx \shownote     \undefined \def \shownote      #1{#1}          \fi
\ifx \showarticletitle \undefined \def \showarticletitle #1{#1}   \fi
\ifx \showURL      \undefined \def \showURL       {\relax}        \fi
% invisible to TeX
\providecommand\bibfield[2]{#2}
\providecommand\bibinfo[2]{#2}
\providecommand\natexlab[1]{#1}
\providecommand\showeprint[2][]{arXiv:#2}

\bibitem[\protect\citeauthoryear{Amazon}{Amazon}{2021}]%
        {TypeAmazonDB}
\bibfield{author}{\bibinfo{person}{Amazon}.} \bibinfo{year}{2021}\natexlab{}.
\newblock \bibinfo{title}{Data Type Supported by Amazon DynamoDB}.
\newblock
\newblock
\urldef\tempurl%
\url{https://docs.aws.amazon.com/amazondynamodb/latest/APIReference/API_Types_Amazon_DynamoDB.html}
\showURL{%
\tempurl}


\bibitem[\protect\citeauthoryear{APL}{APL}{2021}]%
        {apl}
\bibfield{author}{\bibinfo{person}{Amazon APL}.}
  \bibinfo{year}{2021}\natexlab{}.
\newblock \bibinfo{title}{Understand Alexa Presentation Language (APL)}.
\newblock
\newblock
\urldef\tempurl%
\url{https://developer.amazon.com/en-US/docs/alexa/alexa-presentation-language/understand-apl.html}
\showURL{%
\tempurl}


\bibitem[\protect\citeauthoryear{Bajorek}{Bajorek}{2018}]%
        {Bajorek2018}
\bibfield{author}{\bibinfo{person}{Joan~Palmiter Bajorek}.}
  \bibinfo{year}{2018}\natexlab{}.
\newblock \bibinfo{title}{Voice First Versus the Multimodal User Interfaces of
  the Future}.
\newblock
\newblock
\urldef\tempurl%
\url{https://www.uxmatters.com/mt/archives/2018/10/voice-first-versus-the-multimodal-user-interfaces-of-the-future.php}
\showURL{%
\tempurl}


\bibitem[\protect\citeauthoryear{Br{\"u}ggemeier and Lalone}{Br{\"u}ggemeier
  and Lalone}{2019}]%
        {Brggemeier2019}
\bibfield{author}{\bibinfo{person}{Birgit Br{\"u}ggemeier} {and}
  \bibinfo{person}{Philip Lalone}.} \bibinfo{year}{2019}\natexlab{}.
\newblock \showarticletitle{WoS - Open Source Wizard of Oz for Speech Systems}.
  In \bibinfo{booktitle}{\emph{IUI Workshops}}.
\newblock


\bibitem[\protect\citeauthoryear{Cohen, Giangola, and Balogh}{Cohen
  et~al\mbox{.}}{2004}]%
        {Cohen2004}
\bibfield{author}{\bibinfo{person}{Michael~H. Cohen}, \bibinfo{person}{James~P.
  Giangola}, {and} \bibinfo{person}{Jennifer Balogh}.}
  \bibinfo{year}{2004}\natexlab{}.
\newblock \bibinfo{booktitle}{\emph{Voice User Interface Design}}.
\newblock \bibinfo{publisher}{Addison-Wesley}, \bibinfo{address}{Boston, MA}.
\newblock
\showISBNx{0-321-18576-5}


\bibitem[\protect\citeauthoryear{Doherty, Balaskas, and Doherty}{Doherty
  et~al\mbox{.}}{[n.d.]}]%
        {Doherty2020}
\bibfield{author}{\bibinfo{person}{Kevin Doherty}, \bibinfo{person}{Andreas
  Balaskas}, {and} \bibinfo{person}{Gavin Doherty}.}
  \bibinfo{year}{[n.d.]}\natexlab{}.
\newblock \showarticletitle{The Design of Ecological Momentary Assessment
  Technologies}.
\newblock  \bibinfo{volume}{32}, \bibinfo{number}{3}
  (\bibinfo{year}{[n.\,d.]}), \bibinfo{pages}{257--278}.
\newblock
\showISSN{1873-7951}


\bibitem[\protect\citeauthoryear{Ferreira, Goncalves, Kostakos, Barkhuus, and
  Dey}{Ferreira et~al\mbox{.}}{2014}]%
        {Ferreira2014}
\bibfield{author}{\bibinfo{person}{Denzil Ferreira}, \bibinfo{person}{Jorge
  Goncalves}, \bibinfo{person}{Vassilis Kostakos}, \bibinfo{person}{Louise
  Barkhuus}, {and} \bibinfo{person}{Anind~K. Dey}.}
  \bibinfo{year}{2014}\natexlab{}.
\newblock \showarticletitle{Contextual Experience Sampling of Mobile
  Application Micro-Usage}. In \bibinfo{booktitle}{\emph{Proc. MobileHCI '14}}.
  \bibinfo{pages}{91–100}.
\newblock
\showISBNx{9781450330046}


\bibitem[\protect\citeauthoryear{Holland}{Holland}{2021}]%
        {Holland2021}
\bibfield{author}{\bibinfo{person}{Jennifer Holland}.}
  \bibinfo{year}{2021}\natexlab{}.
\newblock \bibinfo{title}{The Challenges \& Advantages of Conversational User
  Interfaces in 2021}.
\newblock
\newblock
\urldef\tempurl%
\url{https://lform.com/blog/post/the-challenges-advantages-of-conversational-user-interfaces-in-2021/}
\showURL{%
\tempurl}


\bibitem[\protect\citeauthoryear{Intille, Haynes, Maniar, Ponnada, and
  Manjourides}{Intille et~al\mbox{.}}{2016}]%
        {Intille2016uEMA}
\bibfield{author}{\bibinfo{person}{Stephen Intille}, \bibinfo{person}{Caitlin
  Haynes}, \bibinfo{person}{Dharam Maniar}, \bibinfo{person}{Aditya Ponnada},
  {and} \bibinfo{person}{Justin Manjourides}.} \bibinfo{year}{2016}\natexlab{}.
\newblock \showarticletitle{$\mu$EMA: Microinteraction-Based Ecological
  Momentary Assessment (EMA) Using a Smartwatch}. In
  \bibinfo{booktitle}{\emph{Proc. UbiComp '16}}. \bibinfo{pages}{1124–1128}.
\newblock
\showISBNx{9781450344616}


\bibitem[\protect\citeauthoryear{Johnson, Mrini, Hogarth, Moore, Nakashole,
  Weibel, and Farcas}{Johnson et~al\mbox{.}}{2020}]%
        {voli}
\bibfield{author}{\bibinfo{person}{Janet Johnson}, \bibinfo{person}{Khalil
  Mrini}, \bibinfo{person}{Michael Hogarth}, \bibinfo{person}{Alison Moore},
  \bibinfo{person}{Nadpa Nakashole}, \bibinfo{person}{Nadir Weibel}, {and}
  \bibinfo{person}{Emilia Farcas}.} \bibinfo{year}{2020}\natexlab{}.
\newblock \showarticletitle{Voice-Based Conversational Agents for Older
  Adults}. In \bibinfo{booktitle}{\emph{Adjunct Proc. CHI '20}}.
\newblock


\bibitem[\protect\citeauthoryear{Maher, Rebar, and Dunton}{Maher
  et~al\mbox{.}}{2018}]%
        {Maher2018}
\bibfield{author}{\bibinfo{person}{Jaclyn~P. Maher}, \bibinfo{person}{Amanda~L.
  Rebar}, {and} \bibinfo{person}{Genevieve~F. Dunton}.}
  \bibinfo{year}{2018}\natexlab{}.
\newblock \showarticletitle{Ecological Momentary Assessment Is a Feasible and
  Valid Methodological Tool to Measure Older Adults' Physical Activity and
  Sedentary Behavior}.
\newblock \bibinfo{journal}{\emph{Frontiers in Psychology}}
  \bibinfo{volume}{9} (\bibinfo{year}{2018}), \bibinfo{pages}{1485}.
\newblock
\showISSN{1664-1078}


\bibitem[\protect\citeauthoryear{Mrini, Chen, Nakashole, Weibel, and
  Farcas}{Mrini et~al\mbox{.}}{2021}]%
        {Mrini2021MedicalQU}
\bibfield{author}{\bibinfo{person}{Khalil Mrini}, \bibinfo{person}{Chen Chen},
  \bibinfo{person}{Ndapa Nakashole}, \bibinfo{person}{Nadir Weibel}, {and}
  \bibinfo{person}{Emilia Farcas}.} \bibinfo{year}{2021}\natexlab{}.
\newblock \showarticletitle{Medical Question Understanding and Answering for
  Older Adults}. In \bibinfo{booktitle}{\emph{3rd SoCal ML \& NLP Symposium}}.
\newblock


\bibitem[\protect\citeauthoryear{Nagel, Hudson, and Abowd}{Nagel
  et~al\mbox{.}}{2004}]%
        {Nagel2004}
\bibfield{author}{\bibinfo{person}{Kristine~S. Nagel},
  \bibinfo{person}{James~M. Hudson}, {and} \bibinfo{person}{Gregory~D. Abowd}.}
  \bibinfo{year}{2004}\natexlab{}.
\newblock \showarticletitle{Predictors of Availability in Home Life
  Context-Mediated Communication}. In \bibinfo{booktitle}{\emph{Proc. CSCW
  '04}}. \bibinfo{pages}{497–506}.
\newblock


\bibitem[\protect\citeauthoryear{Ponnada, Haynes, Maniar, Manjourides, and
  Intille}{Ponnada et~al\mbox{.}}{2017}]%
        {Ponnada2017}
\bibfield{author}{\bibinfo{person}{Aditya Ponnada}, \bibinfo{person}{Caitlin
  Haynes}, \bibinfo{person}{Dharam Maniar}, \bibinfo{person}{Justin
  Manjourides}, {and} \bibinfo{person}{Stephen Intille}.}
  \bibinfo{year}{2017}\natexlab{}.
\newblock \showarticletitle{Microinteraction Ecological Momentary Assessment
  Response Rates: Effect of Microinteractions or the Smartwatch?}
\newblock \bibinfo{journal}{\emph{Proc. ACM Interact. Mob. Wearable Ubiquitous
  Technol.}} \bibinfo{volume}{1}, \bibinfo{number}{3}, Article
  \bibinfo{articleno}{92} (\bibinfo{year}{2017}).
\newblock


\bibitem[\protect\citeauthoryear{Ponnada, Thapa-Chhetry, Manjourides, and
  Intille}{Ponnada et~al\mbox{.}}{2021}]%
        {Ponnada2021}
\bibfield{author}{\bibinfo{person}{Aditya Ponnada}, \bibinfo{person}{Binod
  Thapa-Chhetry}, \bibinfo{person}{Justin Manjourides}, {and}
  \bibinfo{person}{Stephen Intille}.} \bibinfo{year}{2021}\natexlab{}.
\newblock \showarticletitle{Measuring Criterion Validity of Microinteraction
  Ecological Momentary Assessment (Micro-EMA): Exploratory Pilot Study With
  Physical Activity Measurement}.
\newblock \bibinfo{journal}{\emph{JMIR Mhealth Uhealth}} \bibinfo{volume}{9},
  \bibinfo{number}{3} (\bibinfo{date}{10 Mar} \bibinfo{year}{2021}),
  \bibinfo{pages}{e23391}.
\newblock


\bibitem[\protect\citeauthoryear{QuestionPro}{QuestionPro}{2021}]%
        {ConditionalBranching}
\bibfield{author}{\bibinfo{person}{QuestionPro}.}
  \bibinfo{year}{2021}\natexlab{}.
\newblock \bibinfo{title}{What is survey skip logic and branching?}
\newblock
\newblock
\urldef\tempurl%
\url{https://www.questionpro.com/features/branching.html}
\showURL{%
\tempurl}


\bibitem[\protect\citeauthoryear{Sayago, Neves, and Cowan}{Sayago
  et~al\mbox{.}}{2019}]%
        {Sayago2019}
\bibfield{author}{\bibinfo{person}{Sergio Sayago},
  \bibinfo{person}{Barbara~Barbosa Neves}, {and} \bibinfo{person}{Benjamin~R
  Cowan}.} \bibinfo{year}{2019}\natexlab{}.
\newblock \showarticletitle{Voice Assistants and Older People: Some Open
  Issues}. In \bibinfo{booktitle}{\emph{Proc. CUI '19}}. Article
  \bibinfo{articleno}{7}.
\newblock


\bibitem[\protect\citeauthoryear{Stone and Shiffman}{Stone and
  Shiffman}{1994}]%
        {Stone1994}
\bibfield{author}{\bibinfo{person}{Arthur~A Stone} {and} \bibinfo{person}{Saul
  Shiffman}.} \bibinfo{year}{1994}\natexlab{}.
\newblock \showarticletitle{Ecological momentary assessment (EMA) in behavorial
  medicine.}
\newblock \bibinfo{journal}{\emph{Annals of Behavioral Medicine}}
  (\bibinfo{year}{1994}).
\newblock


\bibitem[\protect\citeauthoryear{Sun, Chen, and Zhang}{Sun
  et~al\mbox{.}}{2020}]%
        {Sun2020}
\bibfield{author}{\bibinfo{person}{Ke Sun}, \bibinfo{person}{Chen Chen}, {and}
  \bibinfo{person}{Xinyu Zhang}.} \bibinfo{year}{2020}\natexlab{}.
\newblock \showarticletitle{"Alexa, Stop Spying on Me!": Speech Privacy
  Protection against Voice Assistants}. In \bibinfo{booktitle}{\emph{Proc.
  SenSys '20}}. \bibinfo{pages}{298–311}.
\newblock


\bibitem[\protect\citeauthoryear{SurveyMonkey}{SurveyMonkey}{2021}]%
        {Surveymonkey2021}
\bibfield{author}{\bibinfo{person}{SurveyMonkey}.}
  \bibinfo{year}{2021}\natexlab{}.
\newblock \bibinfo{title}{Make surveys more engaging when you do these 5
  things}.
\newblock
\newblock
\urldef\tempurl%
\url{https://www.surveymonkey.com/curiosity/make-surveys-more-engaging-with-5-things/}
\showURL{%
\tempurl}


\bibitem[\protect\citeauthoryear{VOLI}{VOLI}{2021}]%
        {volisite}
\bibfield{author}{\bibinfo{person}{UCSD VOLI}.}
  \bibinfo{year}{2021}\natexlab{}.
\newblock \bibinfo{title}{VOLI: Voice Assistant for Quality of Life and
  Healthcare Improvement in Aging Populations}.
\newblock
\newblock
\urldef\tempurl%
\url{http://voli.ucsd.edu}
\showURL{%
\tempurl}


\bibitem[\protect\citeauthoryear{Wit.ai}{Wit.ai}{2021}]%
        {witai}
\bibfield{author}{\bibinfo{person}{Wit.ai}.} \bibinfo{year}{2021}\natexlab{}.
\newblock \bibinfo{title}{Building natural language experieces}.
\newblock
\newblock
\urldef\tempurl%
\url{https://wit.ai}
\showURL{%
\tempurl}


\bibitem[\protect\citeauthoryear{Ziegler}{Ziegler}{2018}]%
        {Ziegler2018}
\bibfield{author}{\bibinfo{person}{Josh Ziegler}.}
  \bibinfo{year}{2018}\natexlab{}.
\newblock \bibinfo{title}{Communication is hard - or How our first approach to
  converastion management caused as many problems as it solved.}
\newblock
\newblock
\urldef\tempurl%
\url{https://medium.com/navigating-the-conversation/communication-is-hard-part-2-1b7529398cc2}
\showURL{%
\tempurl}


\end{thebibliography}
